\documentclass[aip,jcp,reprint]{revtex4-1}

\usepackage{amsmath}
\usepackage{amsfonts}
\usepackage{url}
\usepackage{bm}
\usepackage{bbold}
\usepackage{graphics}
\usepackage{paralist}
\usepackage{graphicx}
\usepackage{afterpage}

\newcommand{\bs}[1]{\boldsymbol{#1}}
\newcommand{\bra}{\langle}
\newcommand{\ket}{\rangle}

\def \El {{\textrm{e}}}
\def \Or {{\textrm{o}}}

\begin{document}

\title{Excited-state diffusion Monte Carlo calculations: a simple and efficient two-determinant ansatz}

\author{Nick S. Blunt}
\email{nicksblunt@gmail.com}
\affiliation{University Chemical Laboratory, Lensfield Road, Cambridge, CB2 1EW, United Kingdom}
\author{Eric Neuscamman}
\email{eneuscamman@berkeley.edu}
\affiliation{Department of Chemistry, University of California, Berkeley, California 94720, USA}
\affiliation{Chemical Sciences Division, Lawrence Berkeley National Laboratory, Berkeley, California 94720, USA}

\begin{abstract}
We perform excited-state variational Monte Carlo and diffusion Monte Carlo calculations using a simple and efficient wave function ansatz. This ansatz follows the recent variation-after-response formalism, accurately approximating a configuration interaction singles wave function as a sum of only two non-orthogonal Slater determinants, and further including important orbital relaxation. The ansatz is used to perform diffusion Monte Carlo calculations with large augmented basis sets, comparing to benchmarks from near-exact quantum chemical methods. The significance of orbital optimization in excited-state diffusion Monte Carlo is demonstrated, and the excited-state optimization procedure is discussed in detail. Diffuse excited states in water and formaldehyde are studied, in addition to the formaldimine and benzonitrile molecules.
\end{abstract}

\maketitle

\section{Introduction}
\label{sec:intro}

The accurate calculation of excited-state properties in quantum chemistry is an important problem, often involving significant challenges compared to ground-state cases. Such excited states are usually more multi-reference and diffuse in nature compared to ground state wave functions, and many convenient features of ground-state algorithms do not carry over. Many methods are used; time-dependent density functional theory (TDDFT)\cite{Runge1984} is perhaps the most popular method, while equation-of-motion coupled cluster (EOM-CC)\cite{Emrich1981, Stanton1993, Krylov2008} provides another frequently-used and accurate alternative. Multi-reference methods such as multi-reference perturbation theory\cite{mrpt} (MRPT) and configuration interaction\cite{Szalay2011} (MRCI) are also important options, particularly given the more strongly-correlated nature of excited states.

These methods are powerful, but nonetheless come with limiting factors, and alternative approaches would often be of benefit. TDDFT has well-known errors for certain classes of states and systems\cite{Dreuw2003}, and canonical EOM-CCSD has a scaling of $\mathcal{O}(N^6)$ with system size, $N$. The cost of methods such as MRPT and MRCI is typically higher still. Quantum Monte Carlo (QMC) methods provide a powerful alternative, having the benefit that they are systematically improvable with increasing trial wave function quality and, in their real-space versions, typically have a low scaling of $\sim \mathcal{O}(N^3)$ per sample (or $\mathcal{O}(N^4)$ for a constant total error)\cite{Foulkes2001}. Such real-space QMC approaches also avoid the basis set formalism of methods such as EOM-CC, providing a quite different approach.

Diffusion Monte Carlo (DMC) is a well-established method, with a scaling of $\mathcal{O}(N^3)$ per sample for a typical Slater-Jastrow wave function, and has thus been applied successfully for many decades\cite{Anderson1975, Umrigar1993, Foulkes2001}. Its use for studying excited states, however, is somewhat more limited. It is simple to prove that DMC is exact for excited states, if given a trial wave function with the exact nodal surface. This can be seen by noting that the fixed-node approximation is equivalent to modifying the Hamiltonian by adding an infinite potential barrier at the nodal surface. This simply forces the wave function to be zero on this surface - the exact eigenstate will still be obtained as the solution of this modified Hamiltonian, provided that the trial nodal surface matches the exact one.

Although excited-state DMC calculations have been performed, and in many cases shown to be successful\cite{Williamson1998, Schautz2004_2, Filippi2009, Scemama2018_2}, some studies have demonstrated a particularly strong dependence on the quality of the trial wave function\cite{Schautz2004}. This is troublesome given the difficulty in obtaining appropriate trial wave function ansatz for excited states, and the relative difficulty of optimizing such functions reliably.

We recently introduced an ansatz that we call the finite-difference linear response (FDLR) wave function\cite{Neuscamman2016, Blunt2017}. The FDLR approach allows one to consider a general wave function, and to construct a significantly more flexible ansatz in its linear response space. For a completely general ansatz this would be expensive, increasing the scaling of the approach due to the extra parameter derivatives required (although it should be emphasized that for the linear response of a Slater determinant, as studied in this work, the scaling can be kept at $\mathcal{O}(N^3)$ per sample,\cite{Clark2011, Filippi2016, Assaraf2017} as for a single determinant wave function and FDLR). A finite-difference approximation is an accurate and efficient alternative. In our previous application\cite{Blunt2017}, and in this work, we consider the linear response of a single Slater determinant. The resulting ansatz contains the full flexibility of a configuration interaction singles (CIS) wave function, and in a variational Monte Carlo (VMC) framework it is simple to further perform orbital reoptimization, known to be of crucial importance for excited states, particularly those with charge transfer character\cite{Subotnik2011, Liu2012}. Thus, the full flexibility of CIS with orbital relaxation effects can be achieved in a wave function of only two non-orthogonal Slater determinants.

In our original study, the FDLR wave function was optimized in VMC by the linear method for a range of states, and compared to EOM-CCSD results. This demonstrated the importance of orbital optimization, particularly for excited states. However, the approach was limited in accuracy due to the simple one- and two-body Jastrow factors used, and small basis sets considered. In this article we perform a study using diffusion Monte Carlo and large augmented basis sets, and comparing to accurate benchmarks from selected configuration interaction (SCI) and full configuration interaction quantum Monte Carlo (FCIQMC), and experimental data where available. We have further updated our trial wave function optimization procedure to include recent advances\cite{Shea2017}.

In Section~\ref{sec:theory} we present the theory of the FDLR wave function and the excited-state optimization procedure used. Results are presented in Section~\ref{sec:results}, including a comparison between the FDLR approach and an explicit multi-determinant expansion, before considering applications to water, formaldehyde, formaldimine and benzonitrile.

\section{Theory}
\label{sec:theory}

\subsection{The finite-difference linear response (FDLR) ansatz}
\label{sec:fdlr_theory}

Begin by considering some wave function ansatz, $|\Phi(\bs{X})\ket$, with $\bs{X}$ representing the underlying parameters. In general, a more flexible wave function can be formed by spanning all first parameter derivatives,
\begin{equation}
| \Psi_{\textrm{EOM}} (\bs{X}, \bs{\mu}) \ket = \sum_i \mu_i \Big| \frac{\partial \Phi(\bs{X})}{\partial X_i} \Big\ket.
\label{eq:EOM}
\end{equation}
Here, $| \Psi_{\textrm{EOM}} (\bs{X}, \bs{\mu}) \ket$ is known as an equation-of-motion (EOM) ansatz. Such EOM wave functions are common in excited-state methods, and it would be powerful to routinely use such an ansatz in excited-state VMC. However, wave function optimization by the linear method requires first parameter derivates of the VMC wave function, $| \Psi_{\textrm{EOM}} \ket$, which would then require second parameter derivatives of $|\Phi(\bs{X}) \ket$. For some specific choices of $|\Phi(\bs{X})\ket$, these derivatives may be calculated exactly with $\mathcal{O}(N^3)$ scaling per sample. For example when $|\Phi(\bs{X})\ket$ is equal to a Slater determinant, as will be considered here, approaches such as the table method allow this to be achieved\cite{Clark2011, Filippi2016, Assaraf2017}. However the required derivatives may have a higher scaling for a general $|\Phi(\bs{X}) \ket$.

Instead, the FDLR approach of Ref.~(\onlinecite{Neuscamman2016}) approximates $| \Psi_{\textrm{EOM}} (\bs{X}, \bs{\mu}) \ket$ by
\begin{equation}
| \Psi_{\textrm{FDLR}} (\bs{X}, \bs{\mu}) \ket = | \Phi(\bs{X} + \bs{\mu}) \ket - | \Phi(\bs{X} - \bs{\mu}) \ket,
\end{equation}
correct up to an unimportant normalization factor, provided that $\bs{\mu}$ is scaled to be sufficiently small.

In the initial application\cite{Neuscamman2016} FDLR was applied in orbital space VMC, with $| \Phi \ket$ taken as a Jastrow antisymmetric geminal power (JAGP) wave function. FDLR was then subsequently applied in real space VMC to a Slater determinant ansatz\cite{Blunt2017}, as further considered now.

Define a basis of molecular orbitals, $\{ \phi \}$, constructed as a linear combination of atomic orbitals $\{ \chi \}$,
\begin{equation}
\phi_p(\bs{r}) = \sum_{\mu} \chi_{\mu} (\bs{r}) C_{\mu p}.
\end{equation}
Orbital rotations can be defined in the usual way by
\begin{equation}
\bs{C} = \bs{C}^{0} e^{-\bs{X}},
\label{eq:orb_opt}
\end{equation}
where $\bs{C}^{0}$ defines initial molecular orbitals, and orbital rotations are parameterized by an antisymmetric matrix $\bs{X}$, ensuring that $e^{-\bs{X}}$ is unitary and that $\{ \phi \}$ remain orthonormal.

A Slater determinant is then given by
\begin{align}
D &= |\phi_1 \phi_2 \ldots \phi_{\frac{N}{2}}|, \\
  &= \textrm{det}(\bs{A}),
\label{eq:det_def}
\end{align}
with the Slater matrix $\bs{A}$ defined by
\begin{equation}
A_{ij} = \phi_j(\bs{r}_i).
\end{equation}
In real space Monte Carlo it is most efficient to take a product of separate spin-up and spin-down determinants, so that the final $\Phi(\bs{R}; \bs{X})$ wave function takes the form
\begin{equation}
\Phi(\bs{R}; \bs{X}) = D^{\uparrow}(\bs{R}^{\uparrow}; \bs{X}^{\uparrow}) D^{\downarrow}(\bs{R}^{\downarrow}; \bs{X}^{\downarrow}),
\label{eq:slater_prod}
\end{equation}
with $\bs{R}$ referring to all electron positions collectively, and $\uparrow/\downarrow$ to spin-up/down quantities in the obvious way.

Thus, the FDLR wave function we consider takes the form
\begin{equation}
\Psi_{\textrm{FDLR}} (\bs{R}; \bs{X}, \bs{\mu}) = \Phi(\bs{R}; \bs{X} + \bs{\mu}) - \Phi(\bs{R}; \bs{X} - \bs{\mu}),
\label{eq:final_fdlr}
\end{equation}
with $\Phi$ defined as in Eq.~(\ref{eq:slater_prod}). This FDLR wave function is a simple difference of two non-orthogonal Slater functions, defined at orbital rotation values $\bs{X} + \bs{\mu}$ and $\bs{X} - \bs{\mu}$.

The usefulness of such a wave function form is that it reproduces a configuration interaction singles (CIS) wave function in the limit of small $\bs{\mu}$. This is most easily seen by considering a determinant in second-quantized notation,
\begin{equation}
| D (\bs{X}) \ket = \textrm{exp} ( - \sum_{p > q} X_{pq} \hat{E}^{-}_{pq} ) \: | D_0 \ket,
\end{equation}
where it becomes apparent that an EOM wave function of the form in Eq.~(\ref{eq:EOM}) gives
\begin{equation}
| \Psi_{\textrm{EOM}} (\bs{\mu}, \bs{X} ) \ket = - \sum_{pq} \mu_{pq} \: \hat{E}^{-}_{pq} | D (\bs{X}) \ket,
\end{equation}
with $\hat{E}^{-}_{pq} = \hat{E}_{pq} - \hat{E}_{qp}$ and $\hat{E}_{pq} = \hat{a}_{p \uparrow}^{\dagger} \hat{a}_{q \uparrow} + \hat{a}_{p \downarrow}^{\dagger} \hat{a}_{q \downarrow}$. This is a CIS wave function, where $\bs{\mu}$ are the expansion coefficients.

In an exact CIS expansion, the coefficients $\bs{\mu}$ would be normalized to have an $L^2$-norm of $1$. To apply the finite-difference approximation, we begin with normalized $\bs{\mu}$ coefficients from a prior CIS calculation, and then apply a multiplicative factor of $0.01$ to all coefficients. The accuracy of this has been considered previously, and is considered again in Sec.~(\ref{sec:multidet}). However, it is always easy to check the VMC energy from the first linear method iteration (where the Jastrow factor is unity) to ensure that the CIS energy is obtained within small statistical errors.

Such a CIS wave function will be of limited accuracy by itself. The QMC approach taken here offers multiple avenues to improve towards chemical accuracy and better, while maintaining an $\mathcal{O}(N^3)$ cost per sample. First, the VMC approach allows the excited-state-specific optimization of the orbital rotation parameters, $\bs{X}$, about which the CIS expansion is performed. This has been shown to be important for many excited states, particularly charge transfer states\cite{Subotnik2011, Liu2012}, where the shift in charge can result in errors of multiple eV, when using orbitals optimized for the ground state. Second, it is simple to include Jastrow factors, which account for substantial dynamical correlation. Third, resulting wave functions can be used in diffusion Monte Carlo in a blackbox manner.

For this study, we use spline-based one- and two-body Jastrow factors, as implemented in the QMCPACK code\cite{qmcpack}. The final wave function ansatz therefore takes the form
\begin{equation}
\Psi_{\textrm{Total}}(\bs{R}) = \textrm{J}(\bs{R}) \: \Psi_{\textrm{FDLR}}(\bs{R}),
\label{eq:final_WF}
\end{equation}
where $J(\bs{R})$ denotes both one- and two-body Jastrow factors, and $\Psi_{\textrm{FDLR}}(\bs{R})$ takes the form defined in Eq.~(\ref{eq:final_fdlr}), a simple difference of two non-orthogonal Slater functions.

\subsection{Optimization of excited-state trial wave functions}
\label{sec:opt_theory}

The optimization of a ground-state wave function by VMC is a common task, typically performed by minimizing the energy, $E = \bra \Psi | \hat{H} | \Psi \ket / \bra \Psi | \Psi \ket$, or the variance of the energy,
\begin{equation}
\sigma^2 = \frac{ \bra \Psi | (\hat{H} - E)^2 | \Psi \ket }{ \bra \Psi | \Psi \ket },
\end{equation}
or a weighted sum of the two. This optimization can be performed by a variety of methods, including stochastic reconfiguration\cite{Sorella2001}, the linear method\cite{Umrigar2007, Zhao2016}, steepest descent\cite{Schwarz2017, Sabzevari2018} or Newton-Raphson approaches\cite{Umrigar2005}. In this article, we solely use the linear method.

The optimization of excited states is considerably more challenging. One approach is minimization of $\sigma^2$, which has a minimum of zero at any exact eigenfunction of the Hamiltonian. In many cases this will be successful, but in others can result in convergence to an undesired state, particularly if the initial trial wave function is not sufficiently accurate.

Instead we consider the following target function,
\begin{align}
\Omega(\Psi, \omega) &= \frac{ \bra \Psi | (\omega - \hat{H} ) | \Psi \ket }{ \bra \Psi | (\omega - \hat{H} )^2 | \Psi \ket }, \\
                     &= \frac{ \omega - E }{ (\omega - E)^2 + \sigma^2 }.
\label{eq:target_fn}
\end{align}
This was first considered in Ref.~(\onlinecite{Zhao2016}), where it was proven that $\Omega$ is minimized by the eigenstate $| \Psi \ket$ of the Hamiltonian whose energy is directly above $\omega$. Thus, this gives an approach to directly target a specific interior state, provided a sufficiently accurate $\omega$ can be estimated. In practice, whether or not an optimizer can reach the minimum will depend on the starting point, and so dependence on the initial trial function is not entirely removed. Nonetheless, this dependence is greatly reduced.

However, as discussed in Ref.~(\onlinecite{Shea2017}), the variational principle defined by minimizing $\Omega(\Psi, \omega)$ is not size consistent. That is, if we have a separable system with subsystems $A$ and $B$, and a wave function that can be factorized as $| \Psi_{AB} \ket = | \Psi_A \ket \otimes | \Psi_B \ket$, then $\Omega_{AB} \ne \Omega_A + \Omega_B$. Thus optimizing the combined system will give a different result to optimizing each subsystem individually, in violation of size consistency. Meanwhile, variance minimization \emph{is} size consistent, $\sigma^2_{AB} = \sigma^2_A +\sigma^2_B$. As discussed above, however, variance minimization is not state selective, and so leads to an undesired state being targeted more frequently than minimization of $\Omega(\Psi, \omega)$.

Instead we use the approach described in Ref.~(\onlinecite{Shea2017}), also used recently in Ref.~(\onlinecite{Zhao2018}), which mixes the two variational principles to give a state selective method which achieves size consistency by end of the optimization. To do this, we perform optimization of $\Omega(\Psi, \omega)$ in the initial linear method iterations, helping convergence towards the desired state. Once convergence is achieved, we gradually modify the variational principle to perform variance minimization. This is done by setting
\begin{equation}
\omega = E - \sigma,
\end{equation}
where $E$ and $\sigma$ are the estimates of the energy and standard deviation of the energy for the current trial wave function. It can be seen that this choice leads to
\begin{equation}
\Omega(\Psi, \omega = E - \sigma) = - \frac{1}{2 \sigma},
\end{equation}
so that optimizing $\Omega$ with this choice of $\omega$ is equivalent to variance minimization. Nonetheless, minimizing $\Omega$ in the initial linear method iterations helps ensure convergence to the desired state.

In practice, it is important to not vary $\omega$ between its initial and final values too quickly. Doing so can alter $\Omega(\Psi)$ such that the current wave function is in the basin of convergence for an undesired minimum. Instead, we alter $\omega$ gradually. In this study, we begin varying $\omega$ after $10$ linear method iterations, and allow a further $10$ iterations for $\omega$ to alter to its final value. We then allow $10 - 20$ further iterations for convergence to the new minimum to finalize. We usually find this approach to be very effective. There are, however, some systems where convergence to the desired minimum remains challenging, as will be discussed in the case of formaldehyde.

For the initial wave function, we take the $\bs{\mu}$ parameters to equal the CIS coefficients for the desired state, scaled by a multiplicative factor relative to the normalized coefficients, ensuring an accurate finite-difference approximation. This factor is taken to be $0.01$, as discussed briefly in Section~\ref{sec:multidet}. The initial orbitals are taken as restricted Hartree--Fock orbitals, defined as $\bs{X}=\bs{0}$. We only allow mixing between orbitals belonging to the same irreducible representation of the symmetry point group. We also only allow rotations between occupied and virtual orbitals; we find no noticeable difference by optimizing occupied-occupied and virtual-virtual rotations. We consider only singlet wave functions in this article, and so enforce $\bs{X}^{\uparrow} = \bs{X}^{\downarrow}$ and $\bs{\mu}^{\uparrow} = \bs{\mu}^{\downarrow}$, although triplet symmetry can be enforced with $\bs{\mu}^{\uparrow} = - \bs{\mu}^{\downarrow}$. The Jastrow factor begins optimization from unity (i.e., all Jastrow parameters equal to 0). The initial value of $\omega$ (used for the first $10$ linear method iterations) is taken as $E - \sigma$ from the initial wave function, estimated by a short VMC run.

\subsection{Pseudopotentials and augmented basis sets}
\label{sec:basis}

In this article we use the BFD pseudopotentials (sometimes referred to as effective core potentials, ECPs) of Burkatzki \emph{et al.}\cite{Burkatzki2007} to replace the $1s$ electrons of all non-hydrogen atoms. These pseudopotentials have corresponding valence V$n$Z basis sets which we make use of. We emphasize that pseudopotentials are only applied to non-hydrogen atoms; it is common in QMC for the hydrogen atom to also be treated by a pseudopotential, but this approach is not taken here.

It is well known that excited states often have a very diffuse character, often requiring single-particle basis sets to be augmented with additional functions. While it is true that real space QMC methods are capable or removing some finite basis set error, we have found this alone to be insufficient for high-accuracy excited-state energies in many cases. This is particularly true for the cases of water and formaldehyde, where we study Rydberg excitations. While cusp conditions are present in real-space QMC regardless of the basis set used, a too-small basis will limit the flexibility of the nodal surface.

For all results in Section~\ref{sec:results}, we take the provided basis sets of Burkatzki \emph{et al.}\cite{Burkatzki2007} and add diffuse functions from the singly- and doubly-augmented basis sets of Dunning and coworkers\cite{Dunning1989, Dunning1992}. We refer to these basis sets as aug-BFD-V$n$Z and d-aug-BFD-V$n$Z. For example, for the aug-BFD-VTZ basis used for water in Section~\ref{sec:multidet} we take the provided VTZ basis of Burkatzki \emph{et al.}, and add all additional functions which are contained in the aug-cc-pVTZ but not cc-pVTZ basis set. We find this approach to work very well, reproducing excitation energies from equivalent frozen-core calculations in corresponding augmented ``correlation-consistent'' basis sets within $0.1$ eV in all cases. In order to demonstrate this, we perform comparison results both with and without pseudopotentials throughout Section~\ref{sec:results}.

We note that highly-accurate DMC calculations have recently been performed on excited states of water and formaldehyde by Scemama \emph{et al.},\cite{Scemama2018_2} also using augmented basis sets for DMC calculations. These were performed with extremely accurate trial wave functions, obtained from large selected CI expansions\cite{Scemama2013, Caffarel2016, Scemama2018} and all-electron calculations, and so differ to the results here. However, the results of Ref.~(\onlinecite{Scemama2018_2}) support our conclusions regarding the importance of using augmented basis sets for these particularly diffuse excited states, even in DMC.

\section{Results}
\label{sec:results}

\begin{figure}[t!]
\centering
\includegraphics{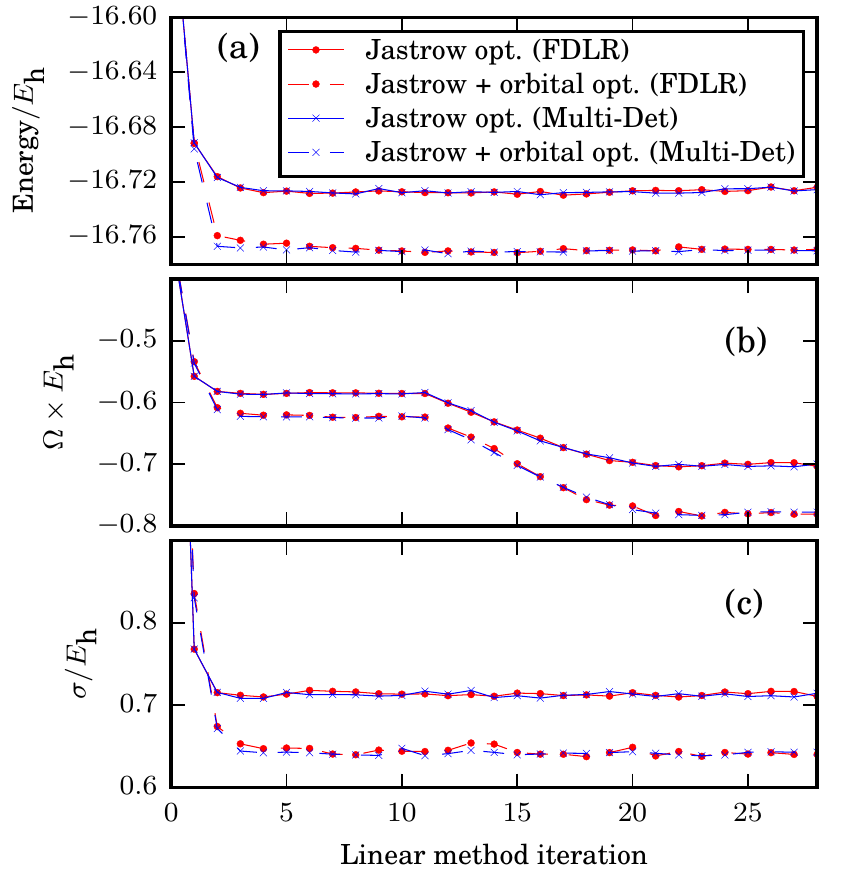}
\caption{Comparison of linear method optimization for FDLR and explicit multi-determinant wave functions. The system is water in an aug-BFD-VTZ basis set (with BFD pseudopotentials), looking at the highly-diffuse $3{}^1\!A_1$ state. Red and blue lines represent FDLR and multi-determinant results, respectively. Solid lines represent Jastrow-only optimization, while dashed lines show simultaneous optimization of the Jastrow and orbitals. Both FDLR and multi-determinant expansions give essentially identical optimizations. (a) shows the local energy profile. (b) shows the profile of $\Omega(\Psi, \omega)$, which is minimized. Between iterations $10$ and $20$, $\omega$ is updated to achieve variance minimization and hence size consistency, as described in the text. (c) shows the standard deviation of the energy, $\sigma$.}
\label{fig:multidet_comp}
\end{figure}

For the following results, all VMC and DMC calculations were performed with the QMCPACK code\cite{qmcpack}. Trial wave functions for QMCPACK were generated using GAMESS\cite{GAMESS}. EOM-CCSD and MRCI benchmarks were obtained using MOLPRO\cite{MOLPRO}. FCIQMC benchmarks were obtained using NECI\cite{NECI_github}. SCI+PT2 (selected CI with second-order perturbation theory) benchmarks used the semi-stochastic heat-bath CI (SHCI) approach, performed using Dice\cite{Dice}. Integral files for NECI and Dice were generated with PySCF\cite{pyscf}, and orbitals visualization were made with IboView\cite{iboview}.

In addition to performing our own benchmarks, we compare to theoretical best estimates (TBE) corrected to the basis set limit from Ref.~(\onlinecite{Loos2018}), where available. This study confirms the importance of highly-augmented basis sets for many of the states considered.

For the wave function optimization we consider optimization of the Jastrow factor alone, denoted $\{J\}$, and optimization of both the Jastrow factor and orbitals simultaneously, $\{J,O\}$. The former case effectively results in a DMC simulation with the nodal surface of a CIS wave function. The ground state wave function is taken to be a single Slater-Jastrow wave function, initialized from Hartree--Fock orbitals. Minimization of $\Omega$ is performed for \emph{both} ground and excited states, ensuring a balanced treatment. We do not consider reoptimization of $\bs{\mu}$ parameters here - we find this typically has a smaller secondary effect compared to orbital reoptimization. It could also lead to some unequal treatment between the ground and excited states, since we include no singly-excited determinants for the ground state ansatz.

DMC calculations were performed with a small time step of $0.001$ au unless stated otherwise. This choice was checked by varying the time step in multiple cases and found to be very accurate. T-moves were used for all systems except water, using the original algorithm of Casula\cite{Casula2006}. T-moves were not found to be necessary for water, however for comparion we also performed all simulations on water \emph{with} T-moves, and present the results in supplementary material. With T-moves enabled, we find absolute energies to increase by $1$ - $2$ m$E_{\textrm{h}}$, but that excitation energies are largely unchanged.

Electron-nuclear and electron-electron Jastrow factors were used, enforcing electron-nuclear and electron-electron cusp conditions (only for hydrogen in the electron-nuclear case, since ECPs were applied for other atoms). These took a spline form with a cutoff at $2a_0$, and with $10$ spline points used for each Jastrow.

Although we focus on DMC excitation energies in this paper, in supplementary material we present absolute energies from both VMC and DMC, $\sigma$ values from VMC, and also present excitation energies from both VMC and DMC in comparison to those from CIS and EOM-CCSD.

Final SCI+PT2 benchmarks were obtained with an extrapolation to a quadratic fit. The error on this extrapolation was estimated in two ways: first as suggested in Ref.~(\onlinecite{Chien2018}), as one fifth of the difference between the best SCI+PT2 estimate and the extrapolated value. Second, using the appropriate element of the covariance matrix of the quadratic fit. We then took whichever was larger. In practice, this uncertainty is only presented if equal to $0.01$ eV or greater.

\subsection{Comparison to a multi-determinant expansion}
\label{sec:multidet}

\begin{table*}
{\footnotesize
\begin{tabular}{ccccccccc}
\hline
\hline
 & & & \multicolumn{5}{c}{Vertical excitation energy/eV} \\
\cline{4-9}
State        &  \;\; ECP/Frozen core? & Basis          & EOM-CCSD     & MRCI-Q   & SHCI    & DMC\{J\} & DMC\{J,O\} & \;\; TBE/Exp. \\
\hline
$2{}^1\!A_1$ & Frozen core            &  aug-cc-pVDZ   & 9.86     & 9.96     & 9.94(1)  &        &          &       \\
             &                        &  aug-cc-pVTZ   & 9.96     & 10.00    & 10.00    &        &          &       \\
             &                        &  aug-cc-pVQZ   & 10.01    & 10.03    & 10.02(1) &        &          &       \\
             &                        & d-aug-cc-pVTZ  & 9.87     & 9.95     & 9.91(1)  &        &          &       \\
             &                        & d-aug-cc-pVQZ  & 9.93     & 9.99     &          &        &          &       \\
\cline{2-8}
             & ECP                    &  aug-BFD-VDZ   & 9.77     & 9.87     & 9.86(1)  & 10.191(8) & 9.973(7)  &       \\
             &                        &  aug-BFD-VTZ   & 9.89     & 9.94     & 9.93(1)  & 10.136(9) & 10.012(7) &       \\
             &                        &  aug-BFD-VQZ   & 9.95     & 9.97     & 9.97(1)  & 10.135(8) & 10.006(7) &       \\
             &                        & d-aug-BFD-VTZ  & 9.82     & 9.90     & 9.85(1)  & 10.071(8) & 9.944(8)  &       \\
             &                        & d-aug-BFD-VQZ  & 9.89     & 9.94     &          &           &           &       \\
\cline{2-8}
             &                        &                &          &          &          &        &          & 9.97\cite{Loos2018}, 9.991\cite{Mota2005}, 9.7\cite{Chutjian1975} \\
\hline
$3{}^1\!A_1$ & Frozen core            &  aug-cc-pVDZ   & 11.76    & 11.86    & 11.84(1) &        &          &       \\
             &                        &  aug-cc-pVTZ   & 11.36    & 11.42    & 11.40    &        &          &       \\
             &                        &  aug-cc-pVQZ   & 11.10    & 11.13    & 11.12(1) &        &          &       \\
             &                        & d-aug-cc-pVTZ  & 10.22    & 10.29    & 10.27(1) &        &          &       \\
             &                        & d-aug-cc-pVQZ  & 10.29    & 10.34    &          &        &          &       \\
\cline{2-8}
             & ECP                    &  aug-BFD-VDZ   & 11.71    & 11.81    & 11.80(2) & 11.23(1)  & 11.075(9)  &       \\
             &                        &  aug-BFD-VTZ   & 11.32    & 11.37    & 11.34(1) & 10.92(1)  & 10.712(10) &       \\
             &                        &  aug-BFD-VQZ   & 11.05    & 11.08    & 11.07(1) & 10.73(1)  & 10.502(9)  &       \\
             &                        & d-aug-BFD-VTZ  & 10.19    & 10.27    & 10.24(1) & 10.390(9) & 10.202(8)  &       \\
             &                        & d-aug-BFD-VQZ  & 10.26    & 10.31    &          &           &            &       \\
\cline{2-8}
             &                        &                &          &          &          &        &          & 10.17\cite{Mota2005,Chutjian1975}, 10.16\cite{Wang1977} \\
\hline
\hline
\end{tabular}
}
\caption{Vertical excitation energies for water, comparing various methods and basis sets. For non-QMC methods, a comparison between frozen-core and pseudopotential results is made, with good agreement generally. MRCI-Q uses a (8\El,11\Or) active space and the Pople correction in Molpro\cite{MOLPRO}, state-averaged over the ground and two excited states studied. SHCI benchmarks are essentially exact within each basis set, and demonstrate that MRCI-Q results are very accurate. These further agree with EOM-CCSD within about 0.1 eV for each basis set. When optimizing only the Jastrow ($\{J\}$), DMC excitation energies are typically too high by $0.2 - 0.3$ eV, which is corrected by orbital reoptimization. Final DMC results also agree well with theoretical best estimates from Ref.~(\onlinecite{Loos2018}), and with experimental values. Note that for the $3{}^1\!A_1$ state, DMC significantly corrects basis set error compared to orbital space methods such as EOM-CC. However, for very good accuracy, doubly-augmented basis sets are required even for DMC.}
\label{tab:water_excit}
\end{table*}

An important consideration is to what extent the finite-difference approximation is capable of reproducing exact CIS results. This was considered in Ref.~(\onlinecite{Blunt2017}) for several excited states of formaldehyde. It was shown that the CIS energy could be reproduced with an accuracy of $1$ m$E_{\textrm{h}}$ for all states considered with a multiplicative factor (applied to initially-normalized CIS coefficients) as large as $0.64$. As described in Section~\ref{sec:results}, we use a factor of $0.01$ for all results in this article, which should be sufficiently small that the finite-difference approximation is negligible.

Here, rather than simply comparing the initial FDLR energy with the exact CIS energy, we perform a linear method optimization of both the FDLR wave function and an explicit multi-determinant CIS expansion in VMC. The multi-determinant expansion is taken to contain exactly the determinants of the CIS wave function. Optimization of orbitals for a multi-determinant expansion has been implemented in QMCPACK\cite{qmcpack}, so that we perform a comparison of both Jastrow-only and Jastrow-orbital optimizations.

We take the water molecule (discussed further in Section~\ref{sec:water}) and consider the highly-diffuse $3{}^1\!A_1$ state. BFD pseudopotentials are used for the oxygen atom, together with the aug-BFD-VTZ basis set. The results of the linear method optimization are presented in Fig.~\ref{fig:multidet_comp}. It can be seen that both the FDLR and multi-determinant ansatz have essentially identical optimization profiles, both for a Jastrow-only optimization, and for simultaneous optimization of orbitals and the Jastrow factor.

The optimization profiles of $\Omega$ and $\sigma$ are also presented in Fig.~\ref{fig:multidet_comp}. This optimization process was described in Section~\ref{sec:opt_theory}. $\Omega(\Psi, \omega)$ is minimized with a constant $\omega$ (taken as $\omega = E - \sigma$ from the initial CIS wave function) until iteration $10$. For iterations $10 - 20$, $\omega$ is gradually updated, and $\Omega$ can be seen to decrease accordingly. As expected, allowing orbital optimization results in lower values of both $\Omega$ and $\sigma$. It also lowers $E$, which is not guaranteed in general, but usually expected for an excited state.

\subsection{Water}
\label{sec:water}

\begin{table*}
\begin{center}
{\footnotesize
\begin{tabular}{@{\extracolsep{4pt}}ccccccc@{}}
\hline
\hline
 & & \multicolumn{5}{c}{Absolute energy/E$_{\textrm{h}}$} \\
\cline{3-7}
State        & Basis          & CCSD     & MRCI-Q     & SHCI       & DMC\{J\}    & DMC\{J,O\}  \\
\hline                             
$1{}^1\!A_1$ &  aug-BFD-VDZ   & -17.184  & -17.191    & -17.1905   & -17.2616(2) & -17.2599(2) \\
             &  aug-BFD-VTZ   & -17.240  & -17.250    & -17.2492   & -17.2624(2) & -17.2611(2) \\
             &  aug-BFD-VQZ   & -17.256  & -17.266    & -17.266    & -17.2623(2) & -17.2606(2) \\
             & d-aug-BFD-VTZ  & -17.240  & -17.251    & -17.2496   & -17.2624(2) & -17.2605(2) \\
             & d-aug-BFD-VQZ  & -17.256  & -17.267    &            &             &             \\
\hline
\hline
\end{tabular}
}
\caption{Absolute energies of the ground state of water, with a BFD pseudopotential applied for the oxygen atom. MRCI-Q uses a (8\El,11\Or) active space and the Pople correction in Molpro\cite{MOLPRO}, state-averaged over the ground and two excited states studied. Orbital optimization $(\{J,O\})$ results in a higher DMC energy compared to the Jastrow-only case $(\{J\})$; this is not unreasonable as the orbitals are reoptimized to minimize variance, while Hartree--Fock orbitals are obtained from energy minimization. T-moves were not used for DMC calculations.}
\label{tab:water_ground}
\end{center}
\end{table*}

We begin by considering the case of water, calculating excitation energies to the $2{}^1\!A_1$ and $3{}^1\!A_1$ states. Excitations of water have been studied extensively, and it is well known that essentially all excitations are highly diffuse, of Rydberg character\cite{Rubio2008}. This makes water an important test case. Although it is a small molecule, this diffuse nature could make it challenging to obtain sufficiently accurate trial states to begin the VMC optimization stage, and potentially make a sufficient initial choice of $\omega$ challenging. We do not use T-moves for DMC here, as they were not found to be necessary.

Vertical excitation energies are presented in Table~\ref{tab:water_excit}. Five basis sets are considered: double-zeta, triple-zeta and quadruple-zeta with single augmentation, and triple-zeta and quadruple-zeta with double augmentation. For the singly-augmented basis sets, we performed accurate benchmarks using MRCI-Q and SHCI. SHCI\cite{Holmes2016_2, Sharma2017} is a type of selected CI method\cite{Huron1973, Buenker1974, Evangelisti1983}, and is essentially exact to the accuracy given. MRCI-Q is less accurate in general, but here we see that it agrees well with SHCI, giving confidence for its use in larger systems and basis sets later, where SHCI (and methods like FCIQMC and DMRG) are more expensive. EOM-CCSD is performed for all basis sets, and is somewhat less accurate, but within $0.1$ eV of SHCI results. Thus we conclude that for this system, EOM-CCSD results in a d-aug-BFD-VTZ basis set, together with MRCI-Q/SHCI results in smaller basis sets, are sensible to compare against. We also compare to experimental results where available, although note that these results are more related to $0$-$0$ transitions than to vertical transitions, so that comparison should only be made cautiously. We further include theoretical best estimates from Ref.~(\onlinecite{Loos2018}), which are more directly comparable.

We also perform non-QMC calculations both with BFD pseudopotentials and with the frozen-core approximation. This gives us a chance to not only test the pseudopotential approximation (usually found to be very good for first row atoms), but also the BFD basis sets used, as adapted here with diffuse functions (see Section~\ref{sec:basis}). We find EOM-CC and MRCI-Q excitation energies to agree very well between the two approaches, always to at least $0.1$ eV accuracy, giving us confidence that the accuracy of the QMC calculations (all of which use pseudopotentials) are not substantially affected by the pseudopotential approximation or basis sets used.

It is seen that, when optimizing only the Jastrow factor (and therefore using the nodal surface of a CIS wave function, represented by FDLR), excitation energies are too high by typically $0.1-0.3$ eV at the basis set limit. Performing orbital optimization corrects almost all of this error, so that DMC results are in good agreement with benchmarks, theoretical best estimates from Ref.~(\onlinecite{Loos2018}), and experimental values.

In Table~\ref{tab:water_ground}, ground state energies are presented for DMC and comparison methods, with a pseudopotential in use for the oxygen again. While ground-state energies are of less interest physically than excitation energies, they give some interesting insight into the VMC and DMC procedure. In particular, it can be seen that DMC energies are actually higher after orbital reoptimization has been performed, compared to the Jastrow-only case. This is perhaps not surprising; optimization is performed by $\Omega$ minimization (and ultimately $\sigma^2$ minimization), while ground-state Hartree--Fock orbitals are obtained by minimizing the ground-state energy (although the final DMC energy only depends on the nodal surface, and so this reasoning is not necessarily correct in all cases). Both $\Omega$ and $\sigma^2$ are lower when reoptimizing orbitals (as required for a correctly performed optimization), and so it is certainly appropriate to favour these results. Moreover, it is important to perform $\Omega$ minimization for \emph{both} ground and excited states for a balanced treatment. Although ground-state energies are higher after orbital optimization, the majority of the reduction in excitation energies comes from a reduction in the excited-state energy. We note that this behavior is not a result of not using T-moves. The same trend is seen if T-moves are used (see supplementary material), and indeed occurs for all systems studied. We also note that this behavior is not a consequence of using a too-large time step. To see this we performed calculations in an aug-BFD-VDZ basis with the original time step of $10^{-3}$ au and then with a time step of $10^{-4}$ au, with T-moves in use. After optimizing only the Jastrow factor, the DMC energy goes from -17.2596(2) $E_{\textrm{h}}$ to -17.2592(4) $E_{\textrm{h}}$ upon reducing the time step. After optimizing both the Jastrow factor and orbitals simultaneously, the DMC energy goes from -17.2582(2) $E_{\textrm{h}}$ to -17.2581(3) $E_{\textrm{h}}$ upon reducing the time step, and so the behavior remains.

\subsection{Testing the optimization procedure}
\label{sec:opt_test}

\begin{figure}[t!]
\centering
\includegraphics{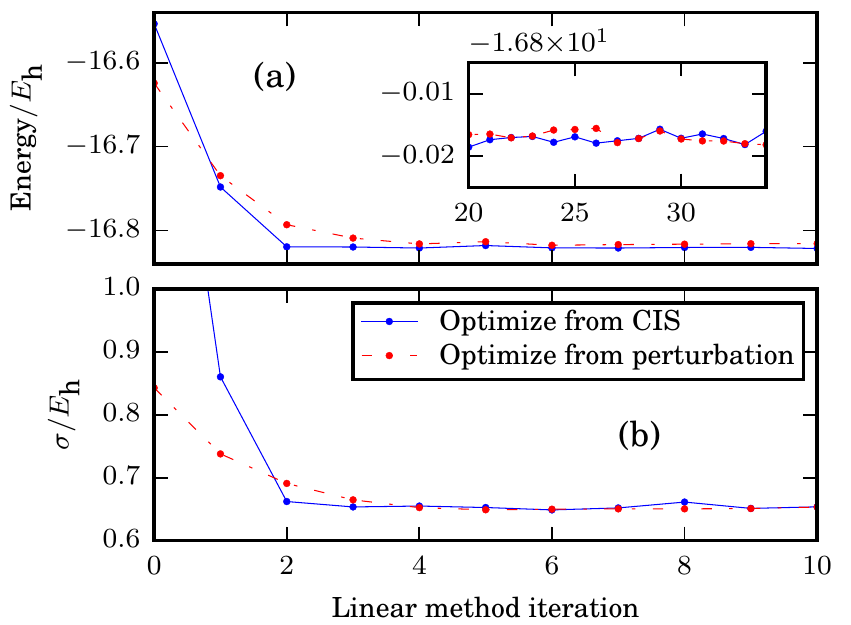}
\caption{Optimization to the $2{}^1\!A_1$ state of water. The solid (red) line shows optimization from a CIS estimate with an initial unity Jastrow factor. For the dashed (blue) line, this final state is taken and perturbed by a significant orbital rotation. The same optimization procedure is then repeated, and obtains the same minimum. The DMC energy after the first optimization (solid line) is -16.8934(2) $E_{\textrm{h}}$. After the perturbation the DMC energy is -17.0205(2) $E_{\textrm{h}}$ - the energy collapses towards the ground state by 127 m$E_{\textrm{h}}$, showing that the nodal surface is severely modified away from that of the excited state. After reoptimization (dashed line) the DMC energy is -16.8935(2) $E_{\textrm{h}}$, showing that the excited-state nodal structure is recovered completely.}
\label{fig:perturb}
\end{figure}

It is interesting to note that, while orbital optimization has a significant effect on the VMC energy (regularly by $30 - 60$ m$E_{\textrm{h}}$, for the systems studied here), the reduction in the DMC energy is relatively small (never much more than $6$ m$E_{\textrm{h}}$ in the same systems). This is somewhat expected - the DMC energy only depends on the nodal surface, rather than whole trial wave function. Nonetheless, it is interesting to test our VMC optimization procedure in a case where the nodal surface is substantially incorrect, and to investigate whether the expected DMC energy can be recovered after reoptimization. This is an important test for the excited-state optimization procedure.

To do so, we once again consider water, using the aug-BFD-VDZ basis and BFD pseudopotential for oxygen, and considering the $2{}^1\!A_1$ state, which has the character of a single HOMO-1 to LUMO excitation from the ground state. First, we take a CIS estimate to this state to form the initial FDLR wave function, and optimize by the procedure described in previous sections. This gives a final optimized wave function, as used in the water results of Section~\ref{sec:water}. We then perform a significant perturbation to substantially modify the nodal surface; this is done by modifying the HOMO-1 to LUMO rotation parameter by -0.5, a large shift. The optimization procedure is then performed again in exactly the same manner.

The optimization profile is shown in Fig.~\ref{fig:perturb}. The optimizer corrects the perturbation in $2$ or $3$ iterations. Although the perturbation does not greatly alter the variance, it has a large effect on energies. In particular, the DMC energy after the initial optimization is calculated as $-16.8934(2) E_{\textrm{h}}$. After the perturbation this falls by 127 m$E_{\textrm{h}}$ to -17.0205(2) $E_{\textrm{h}}$, suggesting that the excited-state nodal structure is destroyed, and significant collapse to the ground state occurs. This is interesting, as the perturbed wave function has a \emph{higher} VMC energy than the optimized excited state, and is thus even farther from the ground state. After the reoptimization the DMC energy is calculated as -16.8935(2) $E_{\textrm{h}}$, fully recovering the original excited-state nodal structure and energy to very good accuracy.

\begin{table*}
\begin{center}
{\footnotesize
\begin{tabular}{@{\extracolsep{4pt}}ccccccccc@{}}
\hline
\hline
 & & \multicolumn{7}{c}{Vertical excitation energy/eV} \\
\cline{3-9}
& & & \;\;\;\; & \multicolumn{2}{c}{\;\;\; DMC, constant $\omega$ \;\;\;} & \multicolumn{2}{c}{\;\;\; DMC, update $\omega$ \;\;\;} \\
\cline{5-6} \cline{7-8}
State        & Basis          & EOM-CCSD & MRCI-Q & \{J\}  & \{J,O\} & \{J\} & \{J,O\} & TBE/Exp. \\
\hline
$2{}^1\!A_1$ &  aug-BFD-VTZ   & 8.20     & 8.22   & 8.59(1) & 8.33(1)  & 8.60(1) & 8.36(1)  &    \\
             &  aug-BFD-VQZ   & 8.25     & 8.25   &         &          &         &          &    \\
             &                &          &        &         &          &         &          & 8.27\cite{Loos2018}, 8.14\cite{Robin1985} \\
\hline
$3{}^1\!A_1$ & d-aug-BFD-VTZ  & 9.33     & 9.44   & 9.70(2) & 9.50(2)  & 9.71(1) & 9.79(1)  &    \\
             & d-aug-BFD-VQZ  & 9.37     & 9.46   &         &          &         &          &    \\
             &                &          &        &         &          &         &          & 9.26\cite{Loos2018} \\
\hline
\hline
\end{tabular}
}
\caption{Vertical excitation energies for formaldehyde. We use different basis sets for the two states, as we find that double augmentation is necessary to properly describe $3{}^1\!A_1$, while single augmentation is sufficient for $2{}^1\!A_1$. DMC calculations are performed with trial wave functions optimized by two schemes. Both schemes perform minimization of $\Omega(\Psi, \omega)$, but in the first case $\omega$ is held constant at its initial value, while in the second case $\omega$ is updated so that $\Omega$ minimization is equivalent to variance minimization. With constant $\omega$, results agree with MRCI-Q values at the basis set limit within  $\approx 0.1$ eV. For the challenging $3{}^1\!A_1$ state, variance minimization results in a higher-energy state being found when orbital optimization is enabled. This is found to be the case even when $\omega$ is updated very slowly. EOM-CCSD and MRCI-Q results are supported by additional extrapolated SHCI results (see main text).}
\label{tab:formaldehyde}
\end{center}
\end{table*}

\subsection{Formaldehyde}
\label{sec:formaldehyde}

Next we consider formaldehyde, another challenging case with many Rydberg excitations, also studied with DMC recently by Loos and coworkers\cite{Scemama2018_2}. Excitations $2{}^1\!A_1$ and $3{}^1\!A_1$ are considered. Calculations performed with EOM-CCSD (varying basis set cardinality and augmentation level) confirm this difficulty; excited states vary greatly between basis sets, to an extent that it is very challenging to clearly identify which state corresponds to which upon changing from single to double augmentation. For example, the $3{}^1\!A_1$ state within singly-augmented basis sets has valence $\pi \rightarrow \pi^*$ character, while the true $3{}^1\!A_1$ state has Rydberg character, as found in a doubly-augmented basis set. This makes this an interesting test for the optimization procedure.

For comparison methods, EOM-CCSD and MRCI-Q, results were calculated in singly- and doubly-augmented basis sets of triple- and quadruple-zeta quality. For the DMC trial function we only consider triple-zeta basis sets, with quadruple-zeta unnecessary. Results are presented in Table~\ref{tab:formaldehyde}. For singly-augmented basis sets, MRCI-Q calculations used a (10\El,11\Or) active space and averaged over the ground and $2{}^1\!A_1$ states. For doubly-augmented basis sets, MRCI-Q calculations used a (10\El,12\Or) active space and averaged over the ground, $2{}^1\!A_1$ and $3{}^1\!A_1$ states. EOM-CCSD compares very well with MRCI-Q for $2{}^1\!A_1$, while giving lower energies by $\approx 0.1$ eV for $3{}^1\!A_1$. We also obtained SCI+PT2 calculations (using SHCI and extrapolations with a quadratic fit) to obtain near-exact benchmarks within the triple-zeta basis sets. For $2{}^1\!A_1$ and aug-BFD-VTZ we obtain an excitation energy of $8.21(3)$ eV, agreeing exactly with MRCI-Q. For $3{}^1\!A_1$ in a d-aug-cc-pVTZ basis the excitation energy is $9.27(3)$ eV [although note that this results uses a frozen core, rather than pseudopotentials, and the geometry of Ref.~(\onlinecite{Loos2018})]\cite{UmrigarEnergy}. This is slightly lower than the excitation energy from MRCI-Q, and in better agreement with EOM-CCSD.

For QMC results we consider two optimization procedures. First as described in Section~\ref{sec:opt_theory}, where $\omega$ is varied to eventually perform $\sigma^2$ minimization for size consistency. We find that this gives difficulties in this case, and so also perform optimization with a fixed $\omega$, using $\omega = E - \sigma$ from the intial trial function. For Jastrow-only optimization, results are too high by $\sim 0.3 - 0.4$ eV compared to best estimates at the basis set limit. Since modifying the Jastrow does not change the nodal surface, the DMC energy only depends slightly on the optimization procedure due to the use of T-moves.

Results are more interesting when orbital optimization is also performed. The $2{}^1\!A_1$ state is relatively simple, and DMC gives good results within $\sim 0.1$ eV of benchmarks for both optimization procedures. The $3{}^1\!A_1$ state is more challenging. When $\omega$ is held constant, results are sensible and as expected. The d-aug-BFD-VTZ excitation energy is $9.50(2)$ eV, comparing reasonably to benchmarks. However, when $\omega$ is updated there is a clear convergence towards a higher-energy state. We have investigated several ways of updating $\omega$ to ensure variance minimization is achieved, but convergence to an undesired minimum always occurs. This is unusual, and highlights the difficulties with such optimizations, particularly towards excited states. It is not clear why this difficulty occurs in this case, although we expect it relates to the challenging nature of states in formaldehyde. It will be interesting in future to explore how much more flexible the ansatz must be made, for example through a large multi-Slater-Jastrow expansion, for the $\Omega$-minimum for the $3{}^1\!A_1$ state to become distinguishable from that of the higher-lying state, as it must eventually be for a sufficiently flexible ansatz.

\subsection{Formaldimine}
\label{sec:formaldimine}

\begin{figure}
\centering
$\; \; \; \; \; \;$
\includegraphics[width=15mm]{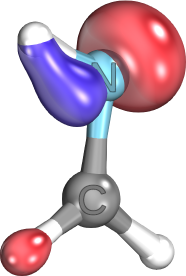}
$\; \; \; \; \; \; \; \; \; \; \;$
\includegraphics[width=40mm]{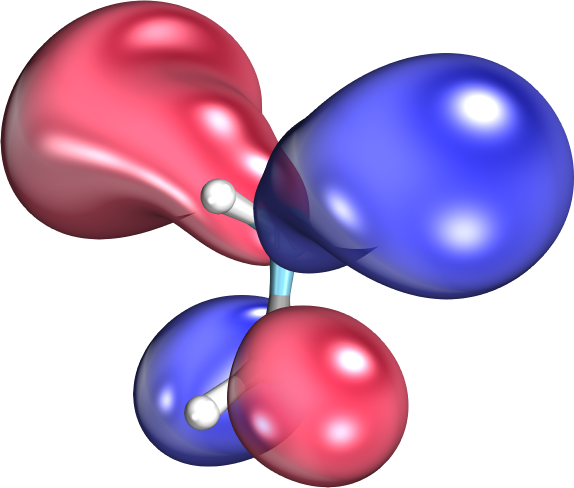}
\caption{Dominant orbitals excited from (left) and to (right) in the studied state of formaldimine. The geometry is also visible, with a small $15$ degree torsional angle applied.}
\label{fig:formaldimine_orbs}
\end{figure}

\begin{table*}
\begin{center}
{\footnotesize
\begin{tabular}{@{\extracolsep{4pt}}cccccccc@{}}
\hline
\hline
 & & & \multicolumn{5}{c}{Vertical excitation energy/eV} \\
\cline{4-8}
State                   & \;\; ECP/Frozen core? & Basis          & EOM-CCSD & MRCI-Q  & DMC\{J\}  & DMC\{J,O\} & TBE \\
\hline
$n \rightarrow \pi^*$   & Frozen core           &  aug-cc-pVDZ   & 5.32     & 5.25    &           &            &     \\
                        &                       &  aug-cc-pVTZ   & 5.29     & 5.21    &           &            &     \\
                        &                       &  aug-cc-pVQZ   & 5.29     & 5.21    &           &            &     \\
\cline{2-7}
                        & ECP                   &  aug-BFD-VDZ   & 5.31     & 5.22    & 5.52(1)   & 5.303(9)   &     \\
                        &                       &  aug-BFD-VTZ   & 5.28     & 5.21    & 5.510(9)  & 5.30(1)    &     \\
                        &                       &  aug-BFD-VQZ   & 5.29     & 5.21    &           &            &     \\
\cline{2-7}
                        &                       &                &          &         &           &            & 5.21\cite{Loos2018} \\
\hline
\hline
\end{tabular}
}
\caption{Vertical excitation energies for formaldimine. DMC energies are too high by $\sim 0.3$ eV compared to MRCI-Q when only the Jastrow factor is optimized. Optimizing the orbitals brings the excitation energy down by around $0.2$ eV, in good agreement with EOM-CCSD, and $\sim 0.1$ eV above MRCI-Q benchmarks. We include EOM-CCSD results in both frozen-core and ECP approximations, showing that the constructed ECP basis sets work well. We further confirmed the accuracy of MRCI-Q by performing i-FCIQMC+PT2\cite{Blunt2018} in the aug-cc-pVDZ basis, with ECPs used, where an excitation energy of $5.26(3)$ eV was obtained, in good agreement with MRCI-Q.}
\label{tab:formaldimine_ver}
\end{center}
\end{table*}

\begin{table*}
\begin{center}
{\footnotesize
\begin{tabular}{@{\extracolsep{4pt}}ccccccc@{}}
\hline
\hline
 & & & \multicolumn{4}{c}{Adiabatic excitation energy/eV} \\
\cline{4-7}
State                   & \;\; ECP/Frozen core? & Basis          & EOM-CCSD & MRCI-Q  &  DMC\{J\} & DMC\{J,O\} \\
\hline                                                                                 
$n \rightarrow \pi^*$   & Frozen core           &  aug-cc-pVDZ   & 4.38     & 4.24    &           &            \\
                        &                       &  aug-cc-pVTZ   & 4.39     & 4.28    &           &            \\
                        &                       &  aug-cc-pVQZ   & 4.41     & 4.29    &           &            \\
\cline{2-7}
                        & ECP                   &  aug-BFD-VDZ   & 4.39     & 4.26    & 4.61(1)   & 4.339(9)   \\
                        &                       &  aug-BFD-VTZ   & 4.42     & 4.30    & 4.63(1)   & 4.314(10)  \\
                        &                       &  aug-BFD-VQZ   & 4.44     & 4.32    &           &            \\
\hline
\hline
\end{tabular}
}
\caption{Adiabatic excitation energies for formaldimine. We also performed i-FCIQMC+PT2 in the aug-BFD-VDZ basis, giving an excitation energy of $4.38(6)$ eV, confirming the accuracy of other benchmark methods within one of two standard errors. Similar to results in Table~\ref{tab:formaldimine_ver}, we find that DMC is too high when optimizing only the Jastrow factor, but that optimizing orbitals gives excitation energies in agreement with MRCI-Q near the basis set limit.}
\label{tab:formaldimine_ad}
\end{center}
\end{table*}

We next consider formaldimine (sometimes called methanimine), CH$_2$NH. This photoactive molecule was studied with DMC by Filippi and coworkers in 2004\cite{Schautz2004}. They found that DMC could be in significant error compared to MRCI depending on the trial wave function used. In particular, the torsional angle (see Fig.~\ref{fig:formaldimine_orbs}) was varied from $0$ to $90$ degrees. Away from $0$ and $90$ degrees the molecule loses its $C_s$ symmetry. The excited state studied is no longer the ground state of a symmetry sector, and DMC becomes more sensitive to the trial wave function quality. Excitation energies were particularly poor at $15$ degrees torsional angle, in error by $\sim 2$ eV for two particular trial wave functions (although these were deliberately basic).

Here we consider the same molecule at the same torsional angle of $15$ degrees, checking that accurate energies can now be obtained. We do not attempt to investigate differences between our results and those of Ref.~(\onlinecite{Schautz2004}) in detail; there are very many differences, including the wave function ansatz used, optimization procedure, basis set, pseudopotentials, among many other subtleties which make such a comparison infeasible. Nonetheless, it serves as an interesting test case.

Both vertical and adiabatic excitation energies to the lowest singlet excited state ($n \rightarrow \pi^*$, see Fig.~\ref{fig:formaldimine_orbs}) are considered. The ground-state geometry was optimized at the CCSD/aug-cc-pVTZ level with the torsional angle held fixed at $15$ degrees. The excited-state geometry for adiabatic excitation energies was optimized similarly with EOM-CCSD/aug-cc-pVTZ with the torsional angle fixed.

Vertical excitation energies are presented in Table~\ref{tab:formaldimine_ver}. EOM-CCSD and MRCI-Q benchmarks were performed from double- to quadruple-zeta quality. We again present these with both frozen-core and pseudopotential approximations, showing that little error is introduced here. MRCI-Q calculations used a (8\El,10\Or) active space. In double-zeta basis sets, we also perform initiator-FCIQMC\cite{Booth2009,Cleland2010}, perturbatively corrected to remove initiator error\cite{Blunt2018}, giving near-exact results to check against. These are presented in the captions of Tables~\ref{tab:formaldimine_ver} and \ref{tab:formaldimine_ad}. EOM-CCSD is higher than MRCI-Q by just under $0.1$ eV in each case.

We only use double- and triple-zeta basis sets for DMC, with quadruple-zeta clearly unnecessary for this state. Without optimizing orbitals, DMC is too high by about $0.3$ eV compared to MRCI-Q, reduced to $0.1$ eV after orbital optimization, with similar accuracy to EOM-CCSD despite the simple trial wave function used. Results are similar for adiabatic energies in Table~\ref{tab:formaldimine_ad}. Performing orbital optimization reduces the DMC excitation energy estimate by $\sim 0.3$ eV, such that results agree with benchmarks to good accuracy.

\subsection{Benzonitrile}
\label{sec:benzonitrile}

\begin{table*}
\begin{center}
{\footnotesize
\begin{tabular}{@{\extracolsep{4pt}}cccccc@{}}
\hline
\hline
 & & & \multicolumn{3}{c}{Vertical excitation energy/eV} \\
\cline{4-6}
State       & \;\; ECP/Frozen core? & Basis          & EOM-CCSD & DMC\{J\} & DMC\{J,O\} \\
\hline
LE          & Frozen core           & aug-cc-pVDZ    & 5.06     &          &            \\
            & ECP                   & aug-BFD-VDZ    & 5.05     & 5.40(5)  & 5.43(4)    \\
\hline
CT          & Frozen core           & aug-cc-pVDZ    & 6.20     &          &            \\
            & ECP                   & aug-BFD-VDZ    & 6.20     & 6.33(3)  & 6.28(4)    \\
\hline
\hline
\end{tabular}
}
\caption{Vertical excitation energies for BN, compared to EOM-CCSD results. The DMC excitation energy for the charge transfer state is within $\sim0.1$ eV of the EOM-CCSD result. However, results are less accurate for the local excitation, differing from EOM-CCSD by $\sim0.4$ eV. Reoptimizing orbitals does not affect results within statistical error bars. This is despite the fact that orbital optimization does reduce the VMC energy and $\sigma$ values by $\sim 25 -35$ m$E_{\textrm{h}}$ for these states (not shown).}
\label{tab:bn}
\end{center}
\end{table*}

As a larger test we consider benzonitrile, investigating two low-lying singlet excited states. Benzonitrile has been quite widely studied\cite{Dixon2015, Medved2015}, in part due to its similarity with the larger 4-aminobenzonitrile (ABN) and 4-dimethylaminobenzonitrile (DMABN)\cite{Grabowski2003}. These molecules share a similar structure in their low-lying excitations: a local excitation (LE) primarily of HOMO-1 to LUMO and HOMO to LUMO+1 character, and a charge transfer (CT) state primarily of HOMO to LUMO character. They exhibit dual fluorescence phenomena involving these states, due to effects believed to be geometric in nature\cite{Grabowski2003, Georgieva2015}. We do not consider such a detailed study here, but take benzonitrile in its ground-state geometry as an example with a low-lying CT state (albeit with a weaker CT character than ABN and DMABN).

We use the ground-state geometry, optimized using CCSD/aug-cc-pVTZ (and the frozen core approximation). It is more challenging to obtain extremely-accurate benchmarks for this system, but EOM-CCSD is sufficient as a comparison method for the accuracy here. Results are presented in Table~\ref{tab:bn}. DMC agrees with EOM-CCSD within $\sim0.1$ eV for the CT state, demonstrating good accuracy. However for the LE state the error relative to EOM-CCSD is $\sim0.4$ eV, which is more significant. Moreover for both states we find that orbital optimization is not significant in improving the DMC energy. At the VMC level, performing orbital optimization does significantly improve each state (including the ground state), reducing each the energy and $\sigma$ value by $\sim 25 - 35$ m$E_{\textrm{h}}$, but clearly this has a much smaller effect on DMC excitation energies. Nonetheless, we do not find that the trial wave function is so poor that collapse towards the ground state occurs. Indeed, DMC tends to overestimate excitation energies. The CT excitation energy being within $\sim 0.1$ eV of EOM-CCSD is an encouraging result, given that our approach has $\mathcal{O}(N^4)$ scaling with system size, compared to $\mathcal{O}(N^6)$ for canonical EOM-CCSD.

It is interesting to consider why the LE excitation energy is relatively poor. One possibility is incomplete optimization; because the linear method involves solving a stochastically-sampled eigenvalue problem, it is possible to observe biases that are only removed in the limit of large sampling\cite{Blunt2018_2}. Sensible approaches to investigate and address this in the future will be through stochastic gradient descent methods\cite{Schwarz2017, Sabzevari2018}. Beyond this, there is much scope for improved trial wave functions, which must eventually remove this error, including the use of advanced Jastrow factors\cite{Umrigar1988, Huang1997, Casula2004, Marchi2009, Goetz2017}.

\section{Conclusion}
\label{sec:conclusion}

We have presented excited-state diffusion Monte Carlo calculations using a finite-difference ansatz consisting of only two non-orthogonal determinants. The optimization of the FDLR ansatz was performed by an approach that is state selective, but which ultimately achieves variance minimization. In the cases of water and formaldimine this was found to be very successful, reproducing accurate calculations at the basis set limit to a typical accuracy of $\sim 0.1$ eV. This is despite the basic nature of the trial wave function, resulting in an algorithm with $\mathcal{O}(N^4)$ scaling.

The optimization procedure was shown to be robust to a perturbation for a very diffuse Rydberg state of water. The optimization was less successful for the $3{}^1\!A_1$ state of formaldehyde, always converging to an undesired minimum upon updating $\omega$ for variance minimization, although sensible results could be achieved if $\omega$ was held constant. This highlights the difficulties of such optimizations in general, which are particularly challenging for excited states. We are investigating alternative approaches to non-linear optimizations that we hope will improve this situation in future work.

The FDLR approach in this paper was applied to the response of a single Slater determinant, equivalent to a CIS ansatz. With modern QMC techniques, such multi-determinant CIS expansions could have been calculated directly with the same scaling achieved here.\cite{Clark2011, Filippi2016, Assaraf2017} However, the FDLR approach should be applicable in a straightforward manner to far more general and flexible wave functions. An example of this was given in Ref.~(\onlinecite{Neuscamman2016}), where the FDLR approach was applied to an antisymmetric geminal power (AGP) ansatz\cite{Casula2004, Marchi2009, Goetz2017, Dupy2018} (in orbital space VMC). Such an AGP wave function has a combinatorial number of determinants when expanded, such that an explicit multi-determinant expansion of the linear response is not a viable option for a polynomial-scaling algorithm. The work presented here makes more realistic the possibility of DMC on a FDLR-AGP wave function, which could be a powerful possibility for accurately treating both static and dynamic correlation in excited states, with polynomial scaling. We also emphasize that the FDLR approach presented here is relatively simple in terms of code. We have implemented FDLR within QMCPACK without significant assumptions on the form of $|\Phi(\bs{X})\ket$, such that generalizing the approach to more flexible wave functions should be straightforward, as should incorporating developments and improved efficiencies\cite{McDaniel2017}. Another possibility for future work is to perform the excited-state orbital optimization deterministically prior to the QMC calculation\cite{Shea2018}. This would remove the most challenging part of our current approach, such that single excitations could be treated simply by DMC for only double the cost of a single-Slater ground-state DMC calculation, and therefore that excited-state simulations up to a thousand electrons may be possible.

\section{Acknowledgments}

This work was supported by the Office of Science, Office of Basic Energy Sciences, the US Department of Energy, under Contract No. DE-AC02-05CH11231.  NSB acknowledges additional support from St John’s College, Cambridge.  QMC calculations on water, formaldehyde and formaldimine were performed on the Berkeley Research Computing Savio cluster.  Larger QMC calculations on benzonitrile were performed at the National Energy Research Scientific Computing Center, a DOE Office of Science User Facility supported by the Office of Science of the U.S. Department of Energy under Contract No. DE-AC02-05CH11231. Dice SHCI calculations were performed on the Cambridge CSD3 Peta4 CPU cluster.

We would like to thank Cyrus Umrigar for providing d-aug-cc-pVTZ SHCI results for formaldehyde via the new Arrow code.

\end{document}